\documentclass[english,12pt]{article}
          \usepackage{babel}
          \usepackage[T1]{fontenc}
          \usepackage[latin2]{inputenc}
          \usepackage{pslatex}
\begin{document}

\title{\bf Trajectory and orbit of the EN200204 {\L}askarzew fireball}
\author{\it Pavel Spurn{\'y}\footnote{Astronomical Institute AV CR,
Fri\v{c}ova 298, 251 65 Ond\v{r}ejov, Czech Republic. Email: {\tt spurny@asu.cas.cz}}, 
~Arkadiusz Olech\footnote{Copernicus Astronomical Center, ul. Bartycka 18,
00-716 Warszawa, Poland. Email: {\tt olech@camk.edu.pl}} ~and~ 
Piotr K\c{e}dzierski\footnote{Warsaw University Observatory, Al. Ujazdowskie
4, 00-478 Warszawa, Poland. Email: {\tt pkedzier@astrouw.edu.pl}}}
\date{Received 2004 March 31}
\maketitle

\begin{abstract}
The fireball of $\sim -10$ mag was observed over Poland on February 20, 2004
at 18:54 UT. Except many visual observations the event was caught by two
photographic stations: one in the Czech Republic and one in Poland. A
description, ground track map, atmospheric trajectory and orbital data
for the fireball are presented.
\end{abstract}

\section{Introduction}

The European Fireball Network (EN) is the project which main goal is to
study properties of meteoroids and their relations to meteorites through
photographic observations of fireballs. The first all-sky cameras,
belonging to EN, were put into operation in Czechoslovakia in 1963. The
number of cameras quickly grew up and now there are about 30 such stations
located in several European countries (Spurn{\'y} 1997).

Polish meteor observers associated in {\it Comets and Meteors Workshop
(CMW)} have many successes in visual and telescopic observations. It is
enough to say, that they collect about 2000 hours of visual observations
and 200 hours of telescopic observations per year sending them regularly
to {\it International Meteor Organization (IMO)} and publishing the
results in {\it WGN} (Olech and Jurek, 2000, Olech, Wi{\'s}niewski and
Gajos, 2001, Z{\l}oczewski, Jurek and Szaruga, 2003) . Unfortunately,
there is still lack of regular photographic meteor observations in
Poland and therefore this country does not belong to EN. It is a serious
problem, taking into account the fact that area of Poland is only
slightly smaller than the area of Germany and four times larger that the
area of the Czech Republic.

To change this situation the CMW decided to buy the photographic and
video cameras with fast lenses in aim of regular monitoring of the sky
over Poland. The details of this project will be published in the
separate contribution to {\it WGN}. The first tests with the new
photographic equipment were made in the end of 2003 and quite regular
observing runs were started in late February of 2004. This paper
presents the results obtained for the $\sim -10$ mag fireball which was
observed on February 20, 2004 over central Poland and was photographed
at Polish station in Ostrowik near Warsaw and EN station no. 16 
Lys\'a hora in the Czech Republic.

\section{Observations}

The EN station  Lys\'a hora uses manually operated all-sky camera with
very precise fish-eye objective Zeiss Distagon 3.5/30 mm. Usually one
exposure per night is taken on the  panchromatic sheet film Ilford FP4
$9\times12$ cm with a sensitivity of 125 ASA. 

The station in Ostrowik uses four Canon T50 cameras equipped with Canon
1.4/50 mm lenses and mounted under the two-arm shutter having frequency
of 5 Hz and producing 10 breaks/sec. The film Konica Centuria 800 ASA,
developed under standard C-41 process, was used. The typical exposure
times were 10-20 minutes. 

\section{The fireball}

The fireball was seen on February 20, 2004 at 18:54 UT by many amateur
astronomers in Poland. The most detailed description comes from
Przemys{\l}aw {\.Z}o{\l}\c{a}dek from Nowy Dw\'or Mazowiecki who saw the
fireball during his telescopic meteors watch. The animation made by him
can be downloaded from: {\tt http://ftp.pkim.org/info/202102bolid.gif}
and the picture of the event caught in Ostrowik is shown in Fig. 1. We
would like to point out that the animation is based only on the visual
observation and in fact the trajectory presented there should be shifted
several degrees to the south and terminated much closer to the horizon.

The fireball traveled its 40.46 km luminous trajectory in 3.22 seconds
and terminated at an altitude of 36.2 km. In fact, it is not real
terminal point because on both stations terminal part of the luminous
trajectory is either out of the field of view (Ostrowik) or behind the
objects on the horizon (Lys{\'a} hora, where the end of the fireball is
behind roof of the station; visible terminal point is there only 5.1
degrees above ideal horizon and 340 km (!) far from the station). So it
is not excluded that the terminal height could be much lower, possibly
below 30 km. It would imply that some smaller part of initial mass of
the order of hundreds of grams in the maximum could survive and land on
the ground. This is also supported by quite high value of the velocity
at the photographic end of the trajectory. It is very probable that the
body could still decelerate to the velocity of some 5 km/s, which could
reach just around 30 km altitude. Then the most probable impact area for
only very small meteorites would lie northward of the city Garwolin and
a little bit south of small village called Puznów Nowy with the center
defined by the following coordinates: $\lambda=21.6461^\circ$ E and
$\phi=51.9095^\circ$ N. However the determination of this impact area is
not very reliable because we have no data about real end of the fireball
luminous trajectory and we do not know real atmospheric profile up to
some 35 km during the fireball flight over this predicted impact area.

The beginning of the fireball was photographed at the height of 71.0 km
over place located about 10 km NE of Kozienice. The maximum brightness 
of around $-10$ mag was reached over {\L}askarzew.  The end of the
photographed trajectory was seen at the height of 36.3 km. The luminous
trajectory of the {\L}askarzew fireball is shown in Fig. 2 and all
important data are collected in Table 1. The orbit of the meteoroid
which caused the EN200204 {\L}askarzew fireball is shown in Fig. 3.

The meteoroid of initial mass of about 2 kg entered the atmosphere with
the velocity of 13.4 km/s and during its detected flight decelerated to a
velocity of 10.0 km/s. The observed radiant of the event is at
$\alpha=90.9^\circ$ and $\delta=+21.4^\circ$.

\begin{table}[h]
\caption{Characteristics of the EN200204 {\L}askarzew fireball}
\begin{center}
\begin{tabular}{lccc}
\hline
\multicolumn{4}{c}{2004 February 20, ${\rm T} = 18^h54^m00^s \pm 20^s$ UT}\\
\hline
\multicolumn{4}{c}{Atmospheric trajectory data}\\
\hline
 & {\bf Beginning} & {\bf Max. light} & {\bf Terminal} \\
Velocity [km/s] & $13.4\pm0.2$ & --- & $10.0\pm0.4$ \\
Height [km] & $71.0\pm0.2$ & --- & $36.3\pm0.2$ \\
Longitude [$^\circ$E] & $21.5874\pm0.0007$ & --- & $21.6266\pm0.0005$\\
Latitude [$^\circ$N] & $51.6324\pm0.0006$ & --- & $51.8130\pm0.0005$\\
Dynamic mass [kg] & 2 & --- & --- \\
Absolute magnitude & $-3$ & $-10^*$ & ---$^*$ \\
Slope [$^\circ$] & $59.59\pm0.04$ & --- & $59.41\pm0.04$ \\
Total length [km] & \multicolumn{3}{c}{40.46}\\
Duration [s] & \multicolumn{3}{c}{3.22}\\
Fireball type & \multicolumn{3}{c}{I or II}\\
Stations & \multicolumn{3}{c}{Lys{\'a} hora, Ostrowik}\\
\hline
\multicolumn{4}{c}{Radiant data (J2000.0)}\\
\hline
 & {\bf Observed} & {\bf Geocentric} & {\bf Heliocentric} \\
Right ascension [$^\circ$] & $90.92\pm0.10$ & $88.50\pm0.13$ & --- \\
Declination [$^\circ$] & $21.40\pm0.10$ & $12.6\pm0.5$ & --- \\
Ecliptical longitude [$^\circ$] & --- & --- & $66.1\pm0.2$ \\
Ecliptical latitude [$^\circ$] & --- & --- & $-2.20\pm0.03$ \\
Initial velocity [km/s] & $13.4\pm0.2$ & $7.5\pm0.4$ & $36.8\pm0.3$\\
\hline
\multicolumn{4}{c}{Orbital data (J2000.0)}\\
\hline
$a$ [AU] & $2.02\pm0.11$ & $\omega$ [$^\circ$] & $13.6\pm0.2$\\
$e$ & $0.52\pm0.03$ & $\Omega$ [$^\circ$] & $151.4310\pm0.0003$\\
$q$ [AU] & $0.9793\pm0.0006$ & $i$ [$^\circ$] & $2.20\pm0.03$\\
$Q$ [AU] & $3.1\pm0.2$ & & \\
\hline
\end{tabular}
\end{center}

\noindent $^*$ - The fireball leaves the FOV of the camera when its
brightness still increases or at least did not fade significantly. It
implies that also in the maximum the fireball could be still brighter.
\end{table}

\bigskip \noindent {\bf Acknowledgments.} ~This work was partially
supported by KBN grant 2~P03D~003~25 to K. Mularczyk.
\bigskip

\noindent {\bf References}

\noindent Spurn{\'y} P. (1997) "Photographic monitoring of fireballs
in Central Europe", SPIE Proceedings, Vol. 3116, p. 144-155

\noindent Olech A. and Jurek M. (2000) "1996-1998 Polish Telescopic
Meteor Database", {\sl WGN}, {\bf 28}, 226-228

\noindent Olech A., Wi{\'s}niewski M. and Gajos M. (2001) "Polish Visual
Meteor Database 1996-1998", {\sl WGN}, {\bf 29}, 214-217

\noindent Z{\l}oczewski K., Jurek M. and Szaruga K. (2003) "Polish Visual
Meteor Database 1999-2001", {\sl WGN}, {\bf 31}, 174-176

\clearpage

~

\vspace{16cm}

\includegraphics{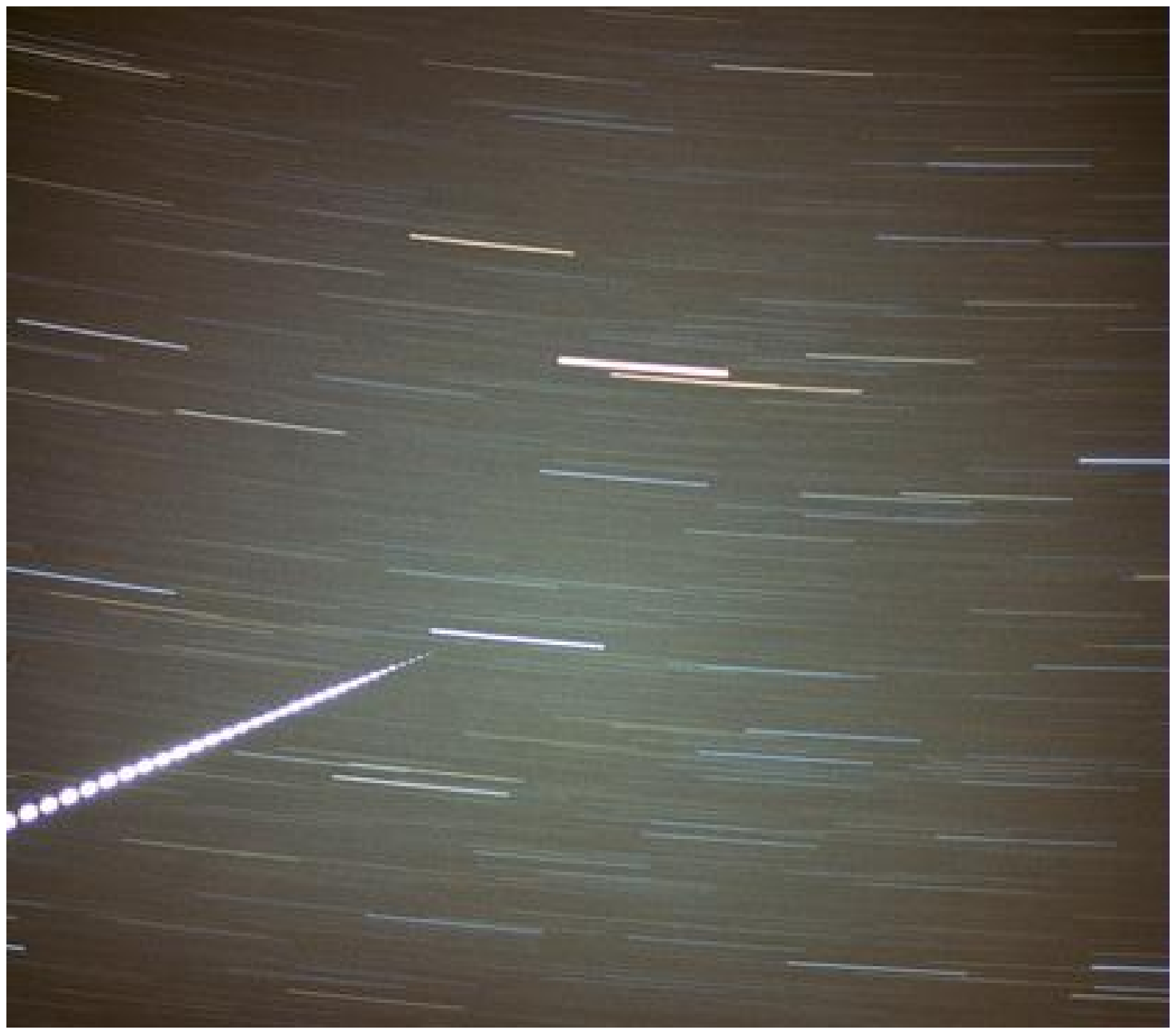}

\noindent {\it Figure 1} - The picture of the the EN200204 {\L}askarzew
fireball taken by Canon T50 camera with Canon 1.4/50 mm lens in Ostrowik
near Warsaw. The brightest object in the picture is planet Saturn.

\clearpage

~

\vspace{15cm}

\includegraphics{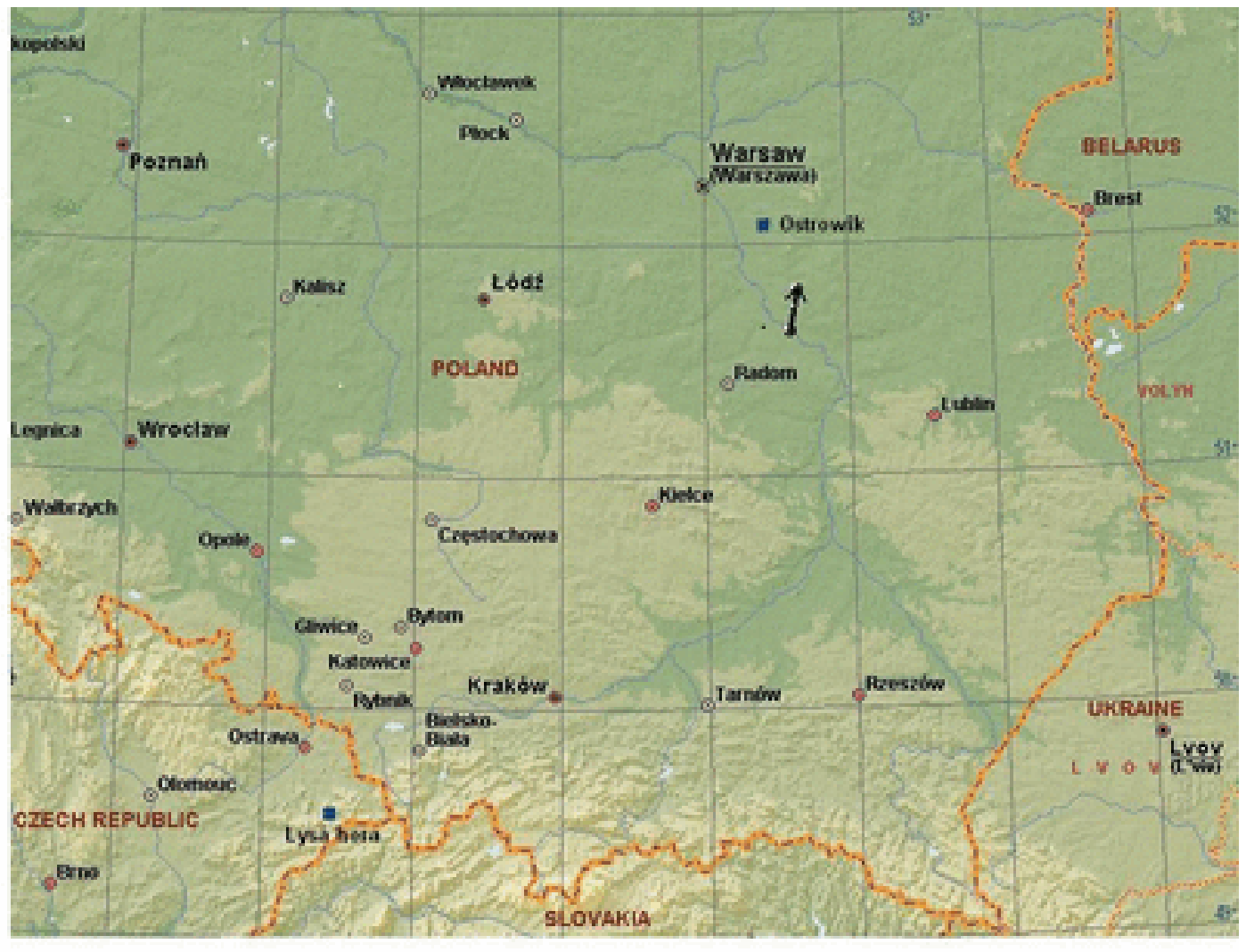}

\noindent {\it Figure 2} - Luminous trajectory of the EN200204
{\L}askarzew fireball.

\clearpage

~

\vspace{16cm}

\includegraphics{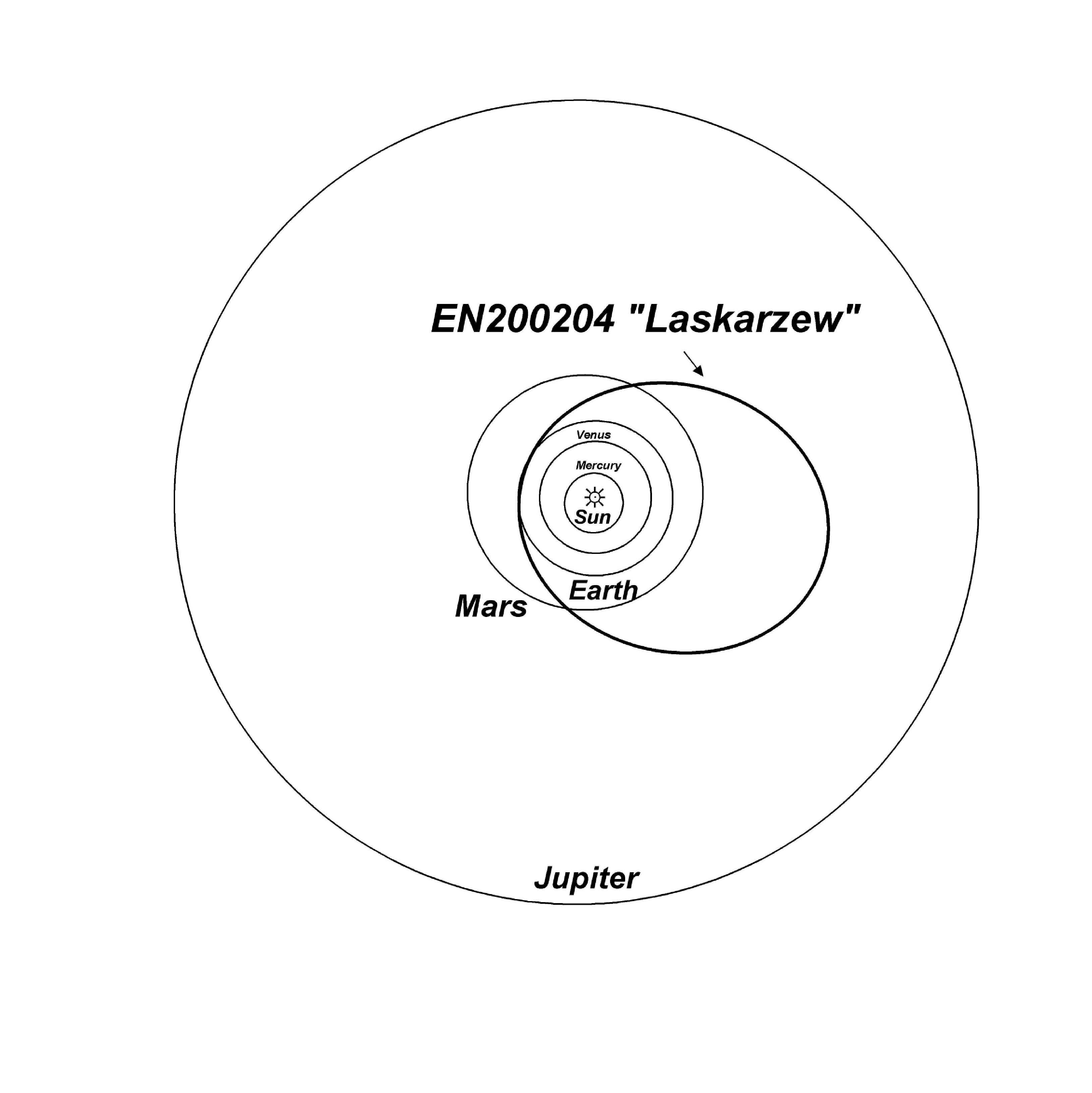}

\noindent {\it Figure 3} - Schematic display of the EN200204
{\L}askarzew fireball orbit projected onto the ecliptic plane.

\end{document}